\documentclass[prd,preprint,floatfix]{revtex4}
\usepackage[dvips]{graphicx}
\usepackage{amssymb,amsmath,bm}

\newcommand{\be}{\begin{equation}}
\newcommand{\ee}{\end{equation}}

\newcommand{\ud}{\text{d}}

\newcommand{\MM}{\mathcal{M}}

\newcommand{\mavgn}[2]{ {\bm\langle} {#2} {\bm\rangle}_{#1} }

\begin{document}

\title{New no-go theorems for \\cosmic acceleration with extra dimensions}

\author{Daniel H. Wesley}
\email{D.H.Wesley@damtp.cam.ac.uk}

\affiliation{Centre for Theoretical Cosmology,
DAMTP, Cambridge University \\
Wilberforce Road, Cambridge CB3 0WA, United Kingdom}

\date{August 4, 2008}
                            
\begin{abstract}
\noindent We describe new no-go theorems for producing four-dimensional accelerating universes from warped dimensional reduction.  
The new theorems improve upon previous results by including dynamical extra dimensions and by treating four-dimensional universes that are not precisely de Sitter.
The theorems show there exists a threshold four-dimensional equation-of-state parameter $w$ below which the number of e-foldings of expansion is bounded, and give expressions for the maximum number of allowed e-foldings. 
In the generic case, the bound must be satisfied if the higher-dimensional theory satisfies the strong energy condition.
If the compactification manifold $\MM$ is one-dimensional, or if its (intrinsic) Ricci scalar $R$ is identically zero, then the bound must be satisfied if the higher-dimensional theory satisfies the null energy condition.
\end{abstract}

\maketitle

\section{Introduction}

Epochs of cosmic acceleration form the keystones of modern cosmological models.  
Observational evidence from type Ia supernovae (SNIa) \cite{SNIa}, the cosmic microwave background (CMB) \cite{Spergel:2006hy}, and other sources indicates that the universe is currently undergoing accelerated expansion, which could be explained by any of a variety of models \cite{Copeland:2006wr}.  Cosmic inflation uses accelerated expansion to account for the near-flatness of the present universe, and to predict a primordial density perturbation spectrum consistent with observations.
  
There are powerful ``no-go'' theorems for accommodating cosmic acceleration in models with extra dimensions \cite{NoGo}. They assume that the strong energy condition (SEC) is obeyed, which requires  \footnote{Our relativity conventions are those of \cite{MTW}.}
\be\label{e:SEC_def}
\left( T_{MN} - \frac{g_{MN}T}{D-2}  \right) t^M t^N \ge 0
\ee
for any non-spacelike vector $t^M$, where $T_{MN}$ the stress-energy tensor in $D$ spacetime dimensions \cite{HE}.  The theorems assert that dimensional reduction on a warped but static internal space  cannot result in a four-dimensional de Sitter universe.  While these are very useful results, the SEC is a very restrictive energy condition and is easily violated: for example by a massive scalar field, or by extended objects such as M or D branes.  The assumption that the universe is undergoing precisely de Sitter acceleration is not a good approximation to present-day observations, though observations are consistent with the universe evolving to a de Sitter phase in the far future.  The extra dimensions could be dynamical, and one might even expect time-dependence in cosmological settings, but the theorems \cite{NoGo} apply only if the extra dimensions are static.  In this letter, we address these issues through improvements of the theorems of \cite{NoGo}.

The first improvement addresses time-dependence, by including dynamical extra dimensions and non-de Sitter four-dimensional universes.  We study universes in which the total four-dimensional energy density $\rho$, including Kaluza-Klein fields and other matter, varies with the Einstein-frame Friedmann-Robertson-Walker (FRW) scale factor $a$ as
\be\label{e:wdef}
\frac{\ud \ln \rho}{\ud \ln a} = -3(1+w).
\ee
with $w$ the equation of state parameter.  We show that including $w > -1$ and dynamical extra dimensions drastically changes the nature of the theorems.  While older results explicitly forbid de Sitter expansion, the new results assert that the number of e-foldings of expansion must be bounded if $w$ falls below a threshold value.  Furthermore, the new theorems give a quantitative formula for the e-folding bound as a function of the number of extra dimensions and the four-dimensional $w$.  As $w \to -1$, the allowed number of e-foldings goes to zero, recovering the previous results of \cite{NoGo} in this limit.

The second improvement is that, in many interesting situations, the energy condition can be weakened from the SEC to the null energy condition (NEC).  The NEC requires that  \cite{HE}
\be\label{eq:NEC_def}
T_{MN} n^M n^N \ge 0 
\ee
for any null vector $n^M$.  It is much more difficult to violate the NEC than the SEC.  Often NEC violation implies pathologies such as superluminal propagation, instabilities, and violations of unitarity.  For two-derivative quantum field theories there are some rigorous formulations of this belief \cite{NEC}.  The NEC can be violated by the Casimir effect, some exotic extended objects (such as orientifold planes or negative-tension branes), and models with higher-derivative actions.  Even in the absence of other pathologies, NEC-violating matter can permit causality violations  \cite{Hawking:1991nk} and a variety of exotic solutions to the Einstein equations which are otherwise forbidden.  Hence it is useful to know under what circumstances one is forced to violate the NEC.

Observationally, the new no-go theorems indicate that experiments with finite resolution in $w$ can discriminate between families of higher-dimensional models that satisfy or violate various energy conditions.  In a purely four-dimensional models, experiments which measure $w$ cannot discriminate between a pure cosmological constant and a scalar field with potential: if the scalar field is in the ``slow-roll" regime, then by flattening its potential the effective $w$ can be made as close to the de Sitter value of $w=-1$ as desired.  Our new no-go theorems show that, in higher-dimensional models, there are thresholds in $w$ which can prevent it from closely approaching to the de Sitter value. Pushing $w$ across these thresholds forces significant changes in the nature of the higher-dimensional theory, and may even be forbidden outright.


\section{Statement of theorems}
 
The no-go theorems depend on certain attributes of the compactification manifold $\MM$, which we take to be closed (without boundary) and compact, or the quotient $\MM'/G$ of a closed compact space $\MM'$ by a discrete group $G$.  In the latter case we work entirely in the ``upstairs" space $\MM'$. We divide the various possibilities for $\MM$  into two categories, which we refer to as ``curvature-free" and  ``curved":

\noindent \emph{-- Curvature-free:}  This category contains any manifold with everywhere vanishing intrinsic Ricci scalar $R$  (more precisely, the Ricci scalar for $f_{ab}$ in (\ref{eq:4kDmetric})).  This is a strong restriction, but it is satisfied by a number of interesting manifolds:  all compact one-dimensional manifolds (in particular compact extra dimensional spaces that appear in braneworld models \cite{Randall:1999ee}),  flat tori, such as those realised as $\mathbb{R}^k/\Lambda$ with $\Lambda$ a lattice, tori with nonnegative Ricci scalar, and compact manifolds of exactly $SU(n)$, $Sp(n)$, $G_2$ and $Spin(7)$ holonomy.  As such it includes the Calabi-Yau complex three-folds and $G_2$ holonomy seven-folds that are essential for realistic four-dimensional compactifications of string and M theory.

\noindent \emph{-- Curved:} This category contains compact manifolds with intrinsic Ricci scalars $R$ that are not identically zero.   Ricci-flat manifolds have no non-Abelian continuous isometries, so many models that realise four-dimensional non-Abelian gauge groups through Kaluza-Klein reduction belong here.  This is essentially the category studied in \cite{NoGo}.

\begin{figure}
  \begin{center}
    \includegraphics[width=3.4in]{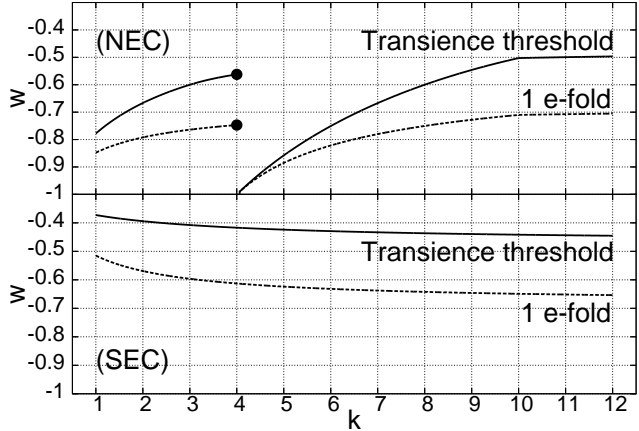}
  \end{center}
  \caption{E-folding bounds for acceleration at constant $w$ for various dimensions $k$ of the compactification manifold $\MM$, for curvature-free $\MM$ (upper panel) and curved $\MM$ (lower panel).  Below the ``Transience threshold" curve, acceleration must be transient, and below the ``1 e-fold" curve, less than one e-folding is allowed.}
  \label{f:efolds}
\end{figure}

The theorems can be stated: \emph{
There is a threshold equation-of-state $w$, and below the threshold a $w$-dependent bound  on the number of e-foldings of expansion.  The bound cannot be violated if the higher-dimensional theory obeys the SEC if $\MM$ is curved, or the NEC if $\MM$ is curvature-free.
}
The e-folding bounds for the simple case where $w$ is constant are illustrated in Figure \ref{f:efolds}.


\section{Additional assumptions}

We have already stated some of the assumptions made about $\MM$: that it be closed and compact (or the quotient of such a space), and fall into the curved or curvature-free categories defined above.  Here we summarize the additional assumptions of the theorems.  We assume the $(4+k)$-dimensional theory is governed by an Einstein-Hilbert action plus matter fields of the form
\be\label{eq:4kDaction}
S_{(4+k)} = \frac{1}{2\kappa^{2+k}} \int \left[
 R(g) + \mathcal{L}_m^{(4+k)} \right] \; \sqrt{-g}\; \ud^{4+k} X
\ee
where $g_{MN}$ is the metric and $X^M$ the coordinates on the full $(4+k)$-dimensional space. $\mathcal{L}_m^{(4+k)}$ is a Lagrangian density for other degrees of freedom, which may depend on $g_{MN}$, but not on its derivatives.  This excludes models, like the DGP model \cite{Dvali:2000hr}, in which the Einstein-Hilbert sector is substantially modified \footnote{This may not be a significant exclusion, since the DGP model suffers from ghosts in the branch most relevant for cosmic acceleration \cite{Koyama:2007za,Charmousis:2006pn}.}.
Nonetheless, models with Ricci terms like $g(\phi)R$ with $\phi$ other fields in the theory, as well as $f(R)$ actions, can be cast in the ``Einstein-frame" form (\ref{eq:4kDaction}) after a suitable conformal transformation.   When we describe violations of the energy conditions, we always mean as interpreted in the Einstein frame.  We use the equations of motion derived from (\ref{eq:4kDaction}) and so enforce a consistent Kaluza-Klein reduction in the sense of \cite{consisKK}. The $(4+k)$-dimensional metric \emph{ansatz} is 
\begin{equation}\label{eq:4kDmetric}
 g_{MN}(t,y)\, \ud X^M \ud X^N  =  \,  e^{2\Omega} \left[ - N(t)^2 \ud t^2 + A(t)^2 \delta_{mn} \ud x^m \ud x^n \right]   + f_{ab}(t,y)\, \ud y^a \ud y^b  
\end{equation}
where $x^m$ and $y^a$ are the coordinates on the three large spatial dimensions and on $\MM$, respectively.  To describe a flat FRW spacetime after dimensional reduction, the metric $f_{ab}(t,y^c)$ and warp function $\Omega(t,y^a)$ on $\MM$ are functions of time $t$ and the extra-dimensional coordinates $y^a$ only.   We parameterise the rate of change of $f_{ab}$  using quantities $\xi$ and $\sigma_{ab}$ defined by
\be
\frac{1}{2} \frac{\ud f_{ab}}{\ud t} = \frac{1}{k} \xi f_{ab} + \sigma_{ab}
\ee
where $f^{ab} \sigma_{ab} =0$.
We place a restriction on the time-evolution of $\MM$, which we discuss below, after introducing the appropriate notation.

One of the most useful tools we apply in the proofs of the theorems is a specific one-parameter family of averages on $\MM$.  Given a parameter $A$ and function $Q(t,y)$ we define the average $\mavgn A Q $ by
\be
\mavgn A Q = \left( \int Q \, e^{A\Omega} \sqrt{f} \; \ud^k y \right)
\left( \int e^{A\Omega} \sqrt{f} \; \ud^k y \right)^{-1}
\ee
hence $\mavgn A Q$ is the average of $Q$ with a volume weighting that depends on the warp factor $\Omega$.
This average resolves $Q(t,y)$ into a $y$-independent mode $Q_{0|A}(t)$ and a ``fluctuation" mode $Q_{\perp|A}(t,y)$ by
\begin{subequations}
\begin{align}
Q_{0|A}(t) &= \mavgn A Q \\
Q_{\perp|A}(t,y) &= Q(t,y) - Q_{0|A}(t)
\end{align}
\end{subequations}
The constant mode $Q_{0|A}$ is the average value of $Q$ by volume, weighted by the warp factor.  For large positive values of $A$ the integration measure is concentrated in regions of $\MM$ with large values of the warp factor $\Omega$.

We define the cosmology experienced by four-dimensional observers by the Einstein frame scale factor $a$ and lapse $n$, not by the five-dimensional quantities $A$ and $N$ appearing in (\ref{eq:4kDmetric}). They are related by
\be\label{e:Credef}
a(t) = A(t)e^{\phi/2} \quad n(t) = N(t)e^{\phi/2}
\ee
where
\be\label{e:PhiDef}
\exp \left( \phi \right) = \int e^{2\Omega} \, f^{1/2} \, \ud^k y = \int \hat f^{1/2} \, \ud^k y
\ee
with $\hat f_{ab} = e^{4\Omega/k} f_{ab}$.  The redefinition (\ref{e:Credef}) is not a conformal transformation of the action, but a change of variables, so no observable quantities are changed. It ensures the four-dimensional action    takes the canonical Einstein-Hilbert form.  This is convenient because cosmological data --  such as SNIa luminosities, CMB temperature maps, and so on -- are usually interpreted assuming Einstein-Hilbert gravity.  Working with Einstein-frame quantities simplifies the comparison with observational evidence.

It is necessary to place a single constraint the time-dependence of $\MM$.  We set
\be\label{eq:GaugeCond}
2\frac{\ud \Omega_{\perp|2}}{\ud t}  + \xi_{\perp|2} = 0
\ee
This condition requires that, if a transformation of $\MM$ changes its volume density as measured by $\hat f^{1/2}$, it is by a ``breathing mode" transformation in which $\MM$ expands or contracts homogeneously.  $\MM$ is allowed to undergo a variety of non-breathing mode transformations, including shear transformations and transformations which change its volume in a $y$-dependent manner as measured in the metric $f_{ab}$.  If the latter occur than (\ref{eq:GaugeCond}) requires a compensating change in the warp factor. Within the moduli space approximation, which is almost universally employed in Kaluza-Klein theory, the manifold $\MM$ evolves adiabatically, and (\ref{eq:GaugeCond}) is a gauge condition.  Without it, dimensional reduction of (\ref{eq:4kDmetric}) yields a scalar sector with kinetic terms of indefinite signature, indicating that within the context of the Kaluza-Klein program  a restriction such as ours is always required \cite{Wesley:2008fg}.


\section{Method of proof}  

To prove the theorems, we use the stress-energy $T_{MN}$ to construct scalar quantities that probe NEC or SEC violation and can be averaged over $\MM$. 
For the NEC we construct
\begin{subequations}\label{e:NECprobes}
\begin{align}
^3N &= - g^{00}T_{00} + (1/3)g^{MN} \, {}^{3}\Pi_M^P  {}^{3}\Pi_N^Q T_{PQ}\\
^kN &= - g^{00}T_{00} + (1/k)g^{MN} \, {}^{k}\Pi_M^P  {}^{k}\Pi_N^Q T_{PQ}
\end{align}
\end{subequations}
where ${}^{3}\Pi_M^P$ and ${}^{k}\Pi_M^P$ are projectors onto the $x^m$ and $y^a$ coordinates, respectively.  That is, ${}^{3}\Pi_M^P = \delta^P_M$ when $M,P$ are coordinate indices on the three large dimensions and vanishes otherwise, while ${}^{k}\Pi_M^P = \delta^P_M$ when $M,P$ are coordinate indices on $\MM$ and vanishes otherwise.  For the SEC, we use the probe $^3N$ and define
\be\label{e:SECprobe}
^kS = -g^{00} T_{00} + \frac{T}{D-2}
\ee
which is obtained by setting $t^M =(1,0,\dots 0)$ in (\ref{e:SEC_def}).  It can be shown that if $^3N$ or $^kN$ are negative at any point, the NEC is violated, and if $^3N$ or $^kS$ are negative at any point, the SEC is violated \cite{Wesley:2008fg}. Because the averaging weight is non-negative, this also implies that, for any $A$, the NEC is violated if the averages $^3N_{0|A}$ or $^kN_{0|A}$ are negative, and the SEC is violated if the averages $^3N_{0|A}$ or $^kS_{0|A}$ are negative.

We illustrate the method of proof for the curvature-free case, for which the relevant energy condition is the NEC.  The proofs for the curved case follow a similar pattern with obvious changes.  When $\MM$ is curvature-free, using the Einstein equations to compute $T_{MN}$ from (\ref{eq:4kDmetric}), constructing $^kN$, and enforcing the NEC by requiring $^kN_{0|A} \ge 0 $ implies 
\be\label{e:masterInEq}
\frac{1}{a^3} \frac{\ud}{\ud t} \left[ a^3 \xi_{0|A} \right]
-
c_0 \xi_{0|A}^2  \ge  c_\perp \mavgn A {\xi^2_{\perp|A}} + 
c_\sigma 
\mavgn A {\sigma^2} + 
 c_\rho \rho + c_\Omega \mavgn A {e^{2\Omega}(\partial \Omega)^2}
\ee
where we have set the lapse $n$ to unity. The coefficients $c_0$, $c_\perp$, $c_\sigma$, $c_\rho$ and $c_\Omega$ are functions of $k$, $w$ and $A$. The $^3N_{0|A} \ge 0$ condition can be written
\be\label{e:masterBC}
\xi_{0|A}^2 \le F(k,w,A,t)
\ee
where $F(k,w,A,t)$ is nonnegative.  Since the NEC implies the SEC, the inequality (\ref{e:masterBC}) applies to both SEC and NEC cases.

The central concept in these proofs is the notion of an ``optimal" solution.
We define the ``optimal" solution at fixed $A$ as the function $\xi_{0|A}$ which solves (\ref{e:masterInEq}) and (\ref{e:masterBC}) for the largest number of e-foldings of $a$.  An optimal solution exists if $c_0 > 0$ and the right-hand side of (\ref{e:masterInEq}) is bounded below as a function of $\xi_{\perp|A}$, $\sigma^2$ and $\Omega$.  This requires all of the coefficients $c_0$, $c_\rho$, $c_\perp$,  $c_\sigma$ and $c_\Omega$ to be positive, or if some coefficients are negative the sum on the right hand side must be positive definite and $c_0 > 0$.  We call an $A$ for which this is true an ``optimising" $A$, and it can be shown  that an optimising $A$ always exists \cite{Wesley:2008fg}.  For an optimising $A$, the optimal solution is the one for which $\xi_{\perp|A}$, $\sigma^2$ and $\Omega$ minimise the right-hand side (typically they are all zero) and for which the inequality (\ref{e:masterInEq}) is saturated.  Any non-optimal solution at $A$ has the same initial conditions as the optimal one, but $\xi_{0|A}$ increases faster, so  (\ref{e:masterBC}) is satisfied for fewer total e-folds.  By studying the optimal solution, we can derive bounds for the simplified choice of $\xi_{\perp|A}$, $\sigma^2$ and $\Omega$, and be guaranteed that any other choice will result in fewer e-foldings of accelerated expansion.

The maximum number of e-foldings is found by choosing an optimising $A$, and saturating (\ref{e:masterInEq}) and (\ref{e:masterBC}).  This gives bounds for arbitrary $w(a)$, but for simplicity we describe the case $w=$ constant.  Defining $v(t) =  t \xi_{0|A}$  yields 
\be\label{e:masterDE}
t \frac{\ud v}{\ud t} = \alpha_2 v^2 + \alpha_1 v + \alpha_0
\ee
from (\ref{e:masterInEq}),
and boundary conditions $v(t_\pm) = \pm v_F$ from (\ref{e:masterBC}), where $t_-$ and $t_+$ are the times at the beginning and end of the accelerating epoch, respectively.  The $\alpha$'s and $v_F$ are all functions of $k$, $w$, and $A$.  For fixed $k$, the right hand side of (\ref{e:masterDE}) vanishes at $v=v^\pm_0$, and the vanishing points $v_0^\pm$ control whether accelerated expansion can be eternal.  If $v^\pm_0 \notin [-v_F,+v_F]$ then $v$ goes from $-v_F$ to $+v_F$ in finite time and acceleration must be transient.  For each $k$ this defines a $w_{\rm thresh}$ such that when $-1 \le w \le w_{\rm thresh}$ acceleration must be transient.  Solving (\ref{e:masterDE}) fixes the ratio $t_+/t_-$ from which  the maximum number of allowed e-folds $N(w)$ can be computed.  The resulting bounds are shown in Figure \ref{f:efolds}.

There is no universal form for the functions $\alpha_0$, $\alpha_1$, $\alpha_2$, $v_F$ and $v^\pm_0$, but these functions vary depending on the choice of extra dimensions and whether $\MM$ is curved or curvature-free.  The full list of functions is given in \cite{Wesley:2008fg}, but here we describe a worked example for $k=1$ extra dimension.  For this example, the functions are
\begin{subequations}
\begin{align}
\alpha_0 &= -\frac{4(1+3w)}{9(1+w)^2} \\
\alpha_1 &= \frac{w-1}{w+1} \\
\alpha_2 &= 1 \\
v_F &= \frac{2}{3}\sqrt{\frac{2}{1+w}} \\
v_0^+ &= \frac{4}{3(1+w)} \\
v_0^- &= - \frac{1+3w}{3(1+w)} 
\end{align}
\end{subequations}
The zero points $v_0^\pm$ lie outside the range $[-v_F,+v_F]$ so long as $w < -7/9$, and so for $w < -7/9$ acceleration must be transient if the five-dimensional theory satisfies the NEC.  Integrating (\ref{e:masterDE}) gives the e-folding bound $N$ with
\be
N(w) = - \frac{4}{5+3w} \, \text{Tanh}^{-1} \left[ \left( \frac{5+3w}{3+5w} \right) \left(\frac{1+w}{2}\right)^{1/2} \right]
\ee
which diverges, as expected, as $w \to -7/9$.  Thus a five-dimensional theory which respects NEC (and obeys the other conditions of the theorems) must give fewer than $N(w)$ e-foldings of acceleration with equation-of-state $w$.

Despite the simplifying assumption that $w$ is constant, the method just described also yields information about for time-varying $w(t)$.  We claim that, if  $w(t) \le w_\star$ with $w_\star$ a constant, then the allowed e-foldings $N[w(t)]$ for time-varying $w(t)$ satisfy $N[w(t)] \le N(w_\star)$.  To prove this, let $\rho$ be the energy density satisfying (\ref{e:wdef}) for time-varying $w(t)$, and $\rho_\star$ be an energy density which matches $\rho$ at a fiducial time $t_0$, and which evolves by (\ref{e:wdef}) with constant $w=w_\star$.  Then (\ref{e:wdef}) implies $\rho \ge \rho_\star$ for $t \ge t_0$.  Since an optimising $A$ exists and $c_\rho$ is nonnegative for this $A$, the right hand side of (\ref{e:masterInEq}) is larger for $\rho$ than it is for $\rho_\star$.  Therefore $\xi_{0|A}$ grows faster and the constraint (\ref{e:masterBC}) is violated after fewer e-folds for $\rho$ than for $\rho_\star$, so $N[w(t)] \le N(w_\star)$.  The techniques described in this letter can be used to obtain refined e-folding bounds for specific functions $w(t)$, by deriving the analogue to  (\ref{e:masterDE}) from (\ref{e:masterInEq}) given the desired $w(t)$.

\section{Conclusions}  

The theorems we described here extend the results of \cite{NoGo} to include a much broader class of theories, by including dynamical extra dimensions and four-dimensional universes that are not precisely de Sitter.  These new theorems indicate that it is possible to escape the conclusions of \cite{NoGo} by introducing time dependence, but that one can do so only temporarily.  The new theorems give a quantitative bound on how closely de Sitter can be approached (in $w$), and for how long transient nearly-de Sitter epochs can last.  In some cases, the new theorems also improve upon the old by weakening the energy condition: for the curvature-free models, the new results show that to obtain nearly-de Sitter acceleration requires violation of the NEC.  This curvature-free class of models is also interesting since it includes the simplest Calabi-Yau and braneworld-type constructions.  Since it is difficult to violate the NEC consistently, it would be quite interesting if we were led to accept the requirement of NEC-violating physics by the hypothesis of extra dimensions and the observation that the universe is accelerating.

On the observational side, these new results place thresholds in $w$, and show how finite experimental resolution can serve to give useful information about potential extra-dimensional physics.  This is especially true for near-future observational surveys which seek to constrain $w$ and its rate of change.  As experimental data improves -- especially data concerned with the present epoch of cosmic acceleration -- the results reported here can be used to rule out large families of models, hopefully giving more clues to the nature of dark energy.

{\bf Acknowledgements.}  We are indebted to Gary Gibbons for useful discussions and for detailed comments.  We thank the Perimeter Institute and the University of Cape Town for their hospitality while completing this work.

\end{document}